\documentclass[prb,twocolumn,superscriptaddress,showpacs,floatfix,amsfonts]{revtex4}
\usepackage{graphicx,graphics,color,epsfig}
\usepackage{bm}
\usepackage{amsmath}
\usepackage{amssymb}
\usepackage{graphicx}

\begin{document}
\preprint{}
\title{Current and Shot Noise in a Quantum Dot Coupled to Ferromagnetic
Leads in the Large $U$ Limit}
\author{Rong L\"{u}}
\affiliation{Center for Advanced Study, Tsinghua University,
Beijing 100084, P. R. China}
\affiliation{Ludwig-Maximilians-Universit\"at, Theresienstr. 37,
               80333 M\"unchen, Germany}
\author{Zhi-Rong Liu}
\affiliation{Max Planck Institute for Metals Research, Heisenbergstrasse 3, 
D-70569 Stuttgart, Germany}
\date{\today}
\begin{abstract}
Using the Keldysh nonequilibrium Green function technique, we study the current and
shot noise spectroscopy of a single interacting quantum dot coupled to two ferromagnetic
leads with different polarizations. The polarizations of leads can be both parallel and antiparallel
alignments. General formulas of current and shot noise are obtained, which
can be applied in both the parallel and antiparallel alignment cases. We show that for large
polarization value, the differential conductance and shot noise are completely diferent
for spin up and spin down configurations in the parallel alignment case. However, the
differential conductance and shot noise have the similar properties for parallel alignment case in the
small polarization value and for antiparallel alignment case in any polarization value. 
\end{abstract}
 \maketitle

Spin-related phenomena in semiconductor quantum dot have attracted great interest recently
as they are the crucial ingredient in the emerging field of spintronics~\cite{Wolf01} and several
quantum computation scheme~\cite{Recher00}. In addition to their potential industrial applications,
these devices also provide an ideal test ground for the study of basic physics including many-body
effect, such as the Kondo effect~\cite{Ng and Glazman88}. 
In this Letter, we use the Keldysh nonequilibrium Green function technique
to study the current and shot noise through an interacting dot coupled to two ferromagnetic (FM) leads
as a function of the applied bias voltage for parallel (P) and antiparallel (AP) lead-polarization alignments.
Using the equation-of-motion approach, we obtain a general formula of current and shot noise
for interacting quantum dot, which can be applied in studying the transport phenomena of dot coupled to
FM leads with both the P and the AP alignments. Our results show that both the differential conductance
and shot noise show the completely different behavior for spin up and spin down configurations in P alignment case 
with large polarization value. However, the differential conductance and shot noise show similar
behavior for P alignment in small polarization value and for AP alignment in any polarization value. 

The system Hamiltonian is written as
\begin{equation}
H=H_{L}+H_{R}+H_{D}+H_{T}\;.
\end{equation}
The Hamiltonian for electrons
in the left and right non-interacting  metallic leads is
\begin{equation}
H_{L}+H_{R}=\sum_{k\in L,R;\sigma} \epsilon_{k\sigma}
c_{k\sigma}^{\dagger}c_{k\sigma}\;,
\end{equation}
where the electron creation (annihilation)
operators in the leads are denoted by $c_{k\sigma}^{\dagger}$ ($c_{k\sigma}$).
The Hamiltonian of the dot is
\begin{equation} H_{D}=\sum_{\sigma}
[\epsilon_{d\sigma}d_{\sigma}^{\dagger}d_{\sigma}
+\frac{U}{2}n_{d,\sigma}n_{d,\bar{\sigma}}]\;,
\end{equation}
where $d_{\alpha}^{\dagger}$ ($d_{\alpha}$) are the creation
(annihilation) operators of dot electrons, and $\epsilon_{d\sigma}
$ is the resonance level of the dot which can be tuned by magnetic field. 
The coupling of the dot to the leads is
\begin{equation}
H_{T}=\sum_{k\in L,R;\sigma,\sigma^{\prime}}
[V_{k\sigma,\sigma^{\prime}}
c_{k\sigma}^{\dagger}d_{\sigma^{\prime}} +H.c.]\;,
\end{equation}
where the tunneling matrix elements $V_{k\sigma,\alpha}$ transfer
electrons through an insulating barrier out of the dot.

Using the
Keldysh nonequilibrium Green function
formalism~\cite{Caroli71,Keldysh65}, the terminal current is given by~\cite{Meir92,Jauho94}
\begin{eqnarray}
I&=&\frac{ie}{4\pi} \int d\epsilon \{
\mbox{Tr}[(f_{L}\mathbf{\Gamma}^{L}-
f_{R}\mathbf{\Gamma}^{R})(\mathbf{G}^{r}
-\mathbf{G}^{a})] \nonumber \\
&&+\mbox{Tr}[ (\mathbf{\Gamma}^{L}-\mathbf{\Gamma}^{R})
\mathbf{G}^{<}]\}\;, \label{EQ:Current1}
\end{eqnarray}
and the spectral density of shot noise in the zero-frequency is given by~\cite{Zhu02}
\begin{eqnarray}
&S(\omega\rightarrow 0)= \frac{e^{2}}{2\pi} \int d\epsilon \{
-f_{L}(1-f_{L})(\mbox{Tr}[(\mathbf{\Gamma}^{L}\mathbf{G}^{r})^{2}]
&\nonumber \\
& +\mbox{Tr}[(\mathbf{\Gamma}^{L}\mathbf{G}^{a})^{2}])
+if_{L}\mbox{Tr}[\mathbf{\Gamma}^{L}\mathbf{G}^{>}]
 -i(1-f_{L})\mbox{Tr}[\mathbf{\Gamma}^{L}\mathbf{G}^{<}]
&\nonumber \\
& +f_{L}\mbox{Tr}[\mathbf{\Gamma}^{L}\mathbf{G}^{>}
\mathbf{\Gamma}^{L}(\mathbf{G}^{r}-\mathbf{G}^{a})]
+\mbox{Tr}[\mathbf{\Gamma}^{L}\mathbf{G}^{>} \mathbf{\Gamma}^{L}
\mathbf{G}^{<}] & \nonumber \\
&-(1-f_{L})
\mbox{Tr}[\mathbf{\Gamma}^{L}(\mathbf{G}^{r}-\mathbf{G}^{a})
\mathbf{\Gamma}^{L}\mathbf{G}^{<}]\}\;,& \label{EQ:Shot1}
\end{eqnarray}
where $f_{L(R)}$ are the Fermi distribution function of the left
and right leads, which has different chemical potential upon a
voltage bias $\mu_{L}-\mu_{R}=eV$. The coupling of the dot to the
leads is characterized by the parameter
\begin{equation}
\Gamma_{\sigma\sigma^{\prime}}^{L(R)}=2\pi\sum_{\sigma^{\prime\prime}}
\rho_{L(R),\sigma^{\prime\prime}}(\epsilon)V_{\sigma^{\prime\prime},\sigma}^{*}(\epsilon)
V_{\sigma^{\prime\prime},\sigma^{\prime}}(\epsilon)\label{EQ:Coupling}
\end{equation}
 with
$\rho_{L(R),\sigma}$ the spin-$\sigma$ band density of states in
the two leads. $G_{\sigma\sigma^{\prime}}^{r(a)}$ and
$G_{\sigma\sigma^{\prime}}^{<(>)}$ are the Fourier transform of the
dot electron  retarded (advanced) Green function
$G_{\sigma\sigma^{\prime}}^{r(a)}(t,t^{\prime})=\mp i\theta(\pm t
\mp t^{\prime})\langle \{d_{\sigma}(t),
d_{\sigma^{\prime}}^{\dagger}(t^{\prime})\}\rangle$,  the lesser
Green function
$G_{\sigma\sigma^{\prime}}^{<}(t,t^{\prime})=i\langle
d_{\sigma^{\prime}}^{\dagger}(t^{\prime})d_{\sigma}(t)\rangle$, and the greater Green function
$G_{\alpha\alpha^{\prime}}^{>}(t,t^{\prime}) =-i\langle
d_{\alpha}(t)d_{\alpha^{\prime}}^{\dagger}(t^{\prime})\rangle$.
It is noted that Eqs. ~(\ref{EQ:Current1}) and ~(\ref{EQ:Shot1}) 
expresses the current and the fluctuations of
current through the quantum dot, an interacting region, in terms
of the distribution functions in the leads and local properties of
the quantum dot, such as the occupation and density of states.

In order to compute the current and the shot noise, one has to 
compute the dot electron retarded Green function $G^{r}$ and
Keldysh Green function $G^{<}$ in the presence of Coulomb interaction $U$.
Without loss of
physics we are considering here,  we assume that the tunneling
matrix elements are spin independent,
$V_{k\sigma,\sigma^{\prime}}=V_{k}
\delta_{\sigma\sigma^{\prime}}$. Using the equation-of-motion
approach, we obtain the retarded dot Green function in the large
$U$ limit as:
\begin{equation}
G_{\sigma}^{r}(\omega)=G_{\sigma\sigma}^{r}(\omega)=\frac{1-\langle
n_{d,\bar{\sigma}}\rangle}{\omega-\epsilon_{d\sigma}-\Sigma_{0\sigma}^{r}
-\Sigma_{1\sigma}^{r}+i0^{+}}\;. \label{EQ:Green-LargeU}
\end{equation}
Here
\begin{eqnarray}
\Sigma_{0\sigma}^{r}(\omega)&=&\sum_{k\in L, R}\frac{\vert
V_{k}\vert^{2}}{\omega-\epsilon_{k\sigma}+i0^{+}} \nonumber \\
&=& \int \frac{d\epsilon}{2\pi}\frac{\Gamma_{\sigma}^{L}(\epsilon)
+\Gamma_{\sigma}^{R}(\epsilon)}{\omega-\epsilon+i0^{+}} \;,
\end{eqnarray}
\begin{eqnarray}
\Sigma_{1\sigma}^{r}(\omega)&=&\sum_{k\in L, R} \frac{\vert
V_{k}\vert^{2}f_{L/R}(\epsilon_{k\bar{\sigma}})}{\omega-\epsilon_{d\sigma}
+\epsilon_{d\bar{\sigma}}
-\epsilon_{k\bar{\sigma}}+i/2\tau_{\bar{\sigma}}}\nonumber \\
 &=& \int
 \frac{d\epsilon}{2\pi}\frac{\Gamma_{\bar{\sigma}}^{L}(\epsilon)f_{L}(\epsilon)
+\Gamma_{\bar{\sigma}}^{R}(\epsilon)f_{R}(\epsilon)}{\omega-\epsilon_{d\sigma}
+\epsilon_{d\bar{\sigma}}- \epsilon+i/2\tau_{\bar{\sigma}}} \;, \label{EQ:Self-energy1}
\end{eqnarray}
where $\Gamma_{\sigma}^{L(R)}(\epsilon)=\Gamma_{\sigma\sigma}^{L(R)}(\epsilon)$,
and the occupation number is subject to the self-consistency
condition $\langle n_{d\sigma}\rangle =-i\int \frac{d\omega}{2\pi}
G_{\sigma}^{<}(\omega)$.
The finite life-time in Eq.~(\ref{EQ:Self-energy1}) for finite
bias voltage and magnetic field can be obtained by using the second-order
perturbation theory~\cite{Meir93}.

The
Green function $G_{\sigma\sigma^{\prime}}^{<(>)}$ cannot be
obtained by directly using the above equation-of-motion approach
without introducing additional assumptions.
By instead applying the operational rules as given by
Langreth~\cite{Langreth76} to the Dyson equation for the
contour-ordered Green function, one can show the following Keldysh
equation for the lesser and greater functions~\cite{Jauho94}
\begin{equation}
\mathbf{G}^{<(>)}(\omega)=\mathbf{G}^{r}(\omega)
\mathbf{\Sigma}^{<(>)}_{T}(\omega)\mathbf{G}^{a}(\omega)\;.
\label{EQ:Green-Lesser}
\end{equation}
From Eq.~(\ref{EQ:Green-LargeU}), the retarded self-energy is
given by
\begin{equation}
\Sigma_{T,\sigma}^{r}(\omega)=\frac{-\langle
n_{d\bar{\sigma}}\rangle(\omega-\epsilon_{d\sigma})
+(\Sigma_{0\sigma}^{r}+\Sigma_{1\sigma}^{r})}{1- \langle
n_{d\bar{\sigma}}\rangle} \;.
\end{equation}
We then arrive at
\begin{eqnarray}
&\Sigma_{T,\sigma}^{<}(\omega)-\Sigma_{T,\sigma}^{>}(\omega)
=\Sigma_{T,\sigma}^{a}(\omega)-\Sigma_{T,\sigma}^{r}(\omega)\nonumber \\
&=i(1-\langle n_{d\bar{\sigma}}\rangle)^{-1}
\{(\Gamma_{\sigma}^{L}(\omega) +\Gamma_{\sigma}^{R}(\omega))&
\nonumber \\
&-2 \mathop{\rm Im}%
\Sigma _{1\sigma }^r\left( \omega \right)\}&\;.
\end{eqnarray}
To determine $\Sigma_{T,\sigma}^{<}$ and $\Sigma_{T,\sigma}^{>}$,
we follow the ansatz proposed in Ref.~\cite{Ng96} to assume
further that these self-energies have the form
\begin{subequations}
\begin{equation}
\Sigma_{T,\sigma}^{<}=i[\Gamma_{\sigma}^{L}(\omega)f_{L}(\omega) +
\Gamma_{\sigma}^{R}(\omega)f_{R}(\omega)]R_{\sigma}(\omega)\;,
\label{EQ:Scatter-IN}
\end{equation}
\begin{equation}
\Sigma_{T,\sigma}^{>}=-i[\Gamma_{\sigma}^{L}(\omega)(1-f_{L}(\omega))
+
\Gamma_{\sigma}^{R}(\omega)(1-f_{R}(\omega))]R_{\sigma}(\omega)\;,
\label{EQ:Scatter-OUT}
\end{equation}
\end{subequations}
where $R(\omega)$ can be regarded as a renormalization factor due
to strong electron-electron Coulomb repulsion in the dot.  A
little algebra yields
\begin{equation}
R_{\sigma}(\omega)=\frac{1}{1-\langle n_{d\bar{\sigma}}\rangle}
\{ 1-\frac{2}{\Gamma_{\sigma}} \mathop{\rm Im}%
\Sigma _{1\sigma }^r\left( \omega \right) \}
\end{equation}
where $\Gamma_{\sigma}=\Gamma_{\sigma}^{L}+\Gamma_{\sigma}^{R}$.

As an extension, we can assume in general the ``scattering in''
and ``scattering out'' self-energies to have the form
\begin{subequations}
\begin{equation}
\mathbf{\Sigma}_{T}^{<}=i[\mathbf{\Gamma}^{L}(\omega)f_{L}(\omega)
+ \mathbf{\Gamma}^{R}(\omega)f_{R}(\omega)]\mathbf{R}(\omega)\;,
\label{EQ:Scatter-IN1}
\end{equation}
\begin{equation}
\mathbf{\Sigma}_{T}^{>}=-i[\mathbf{\Gamma}^{L}(\omega)(1-f_{L}(\omega))
+
\mathbf{\Gamma}^{R}(\omega)(1-f_{R}(\omega))]\mathbf{R}(\omega)\;,
\label{EQ:Scatter-OUT1}
\end{equation}
\label{EQ:Scatter}
\end{subequations}
Eqs.~(\ref{EQ:Scatter-IN1}) and (\ref{EQ:Scatter-OUT1}) leads to
\begin{eqnarray}
\mathbf{G}^{r}-\mathbf{G}^{a}&=&\mathbf{G}^{>}-\mathbf{G}^{<}
\nonumber \\
&=&
-i\mathbf{G}^{r}(\mathbf{\Gamma}^{L}+\mathbf{\Gamma}^{R})\mathbf{R}
\mathbf{G}^{a}\;.\label{EQ:Green-ra}
\end{eqnarray}
Substitution of Eq.~(\ref{EQ:Green-ra}) and
Eq.~(\ref{EQ:Green-Lesser}) with Eq.~(\ref{EQ:Scatter}) into
Eqs.~(\ref{EQ:Current1}) and (\ref{EQ:Shot1}) gives rise to
\begin{eqnarray}
I&=&\frac{e}{2\pi}\int d\epsilon \mbox{Tr}\{ \mathbf{\Gamma}^{L}
[\mathbf{G}^{<} + f_{L}(\mathbf{G}^{r}-\mathbf{G}^{a})]\}\nonumber
\\
&=&\frac{e}{2\pi}\int d\epsilon \mbox{Tr}\{ \mathbf{\Gamma}^{L}
[i\mathbf{G}^{r}(f_{L}\mathbf{\Gamma}^{L}+f_{R}\mathbf{\Gamma}^{R})
\mathbf{R}\mathbf{G}^{a}\nonumber \\
&& -i f_{L}\mathbf{G}^{r}(\mathbf{\Gamma}^{L}+\mathbf{\Gamma}^{R})
\mathbf{R}\mathbf{G}^{a}]\}\nonumber
\\
&=&\frac{e}{2\pi}\int d\epsilon [f_{L}-f_{R}]
\mbox{Tr}[\mathbf{T}]\;, \label{EQ:Current2}
\end{eqnarray}
and
\begin{eqnarray}
S&=& \frac{e^{2}}{2\pi} \int d\epsilon \{ [f_{L}(1-f_{L})
\mbox{Tr}\{\mathbf{T}[1+2(\mathbf{\Gamma}^{R}\mathbf{R})^{-1}(\mathbf{R}-1)]\}
\nonumber \\
&&+f_{R}(1-f_{R})\mbox{Tr}[\mathbf{T}]
 +(f_{L}-f_{R})^{2} \mbox{Tr}[(1-\mathbf{T})\mathbf{T}]\}\;.
\label{EQ:Shot2}
\end{eqnarray}
where the transmission coefficient matrix is defined as
$\mathbf{T}=\mathbf{G}^{a}\mathbf{\Gamma}^{L} \mathbf{G}^{r}
\mathbf{\Gamma}^{R}\mathbf{R}$. We remark that Eqs. ~(\ref{EQ:Current2}) and
~(\ref{EQ:Shot2}) are the general formulas for current and shot noise in 
the interacting dots with large $U$,
which can be applied in studying the spin-transport of both the P and AP configurations.
It is noted that the formula of proportional coupling (i. e., $\Gamma^{L}_{\sigma}=\lambda \Gamma^{R}_{\sigma}$)
does not work in the AP alignment case at finite bias voltage and magnetic field. 
However, one can easily to compute the current and shot noise
for AP alignment configuration by using our formalisms
Eqs. ~(\ref{EQ:Current2}) and
~(\ref{EQ:Shot2}).
It is easy to find that for the
noninteracting case, the above two expressions are just
the Landauer-B\"{u}ttiker formalisms developed based on the scattering matrix
theory. The connection between the two formalisms
was first established by Meir and Wingreen for the current~\cite{Meir92},
and by Zhu and Balatsky for the shot noise~\cite{Zhu02}.

\begin{figure}
\includegraphics[width=8.5cm]{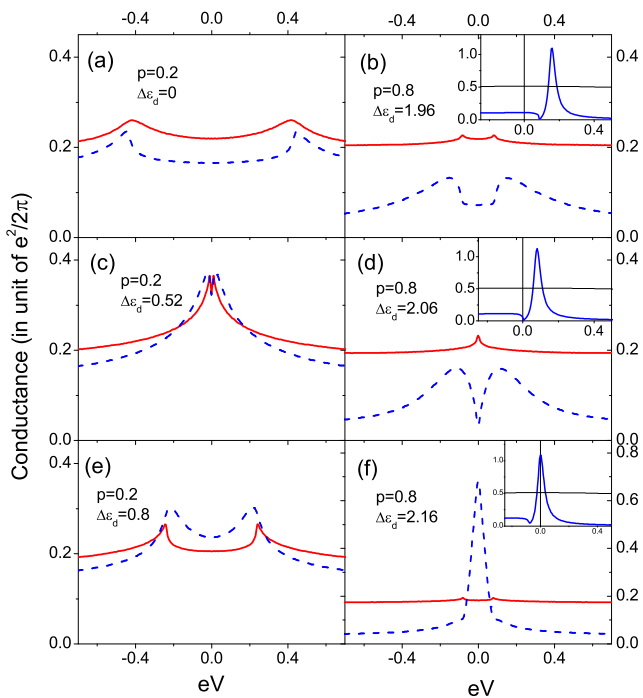}
\caption{\label{fig01} The zero-temperature differential
conductance of spin up (red-solid line) and spin down (blue-dashed line)
for FM leads with P alignment as a function of bias voltage without (a) and
with (b-f) Zeeman splitting $\Delta\epsilon_{d}=\epsilon_{d\uparrow}-\epsilon_{d\downarrow}$.
Here $\Gamma_{0}=1$, the dot level in the absence of magnetic field is
$\epsilon_{d\uparrow}=\epsilon_{d\downarrow}=-3.0$, and the band width is 100. Insets in
(b,d,f) indicate the DOS of spin down configuration as a function of bias voltage near $eV=0$.}
\end{figure}

In the wideband limit, we assume that ${\Gamma}^{L(R)}_{\sigma}$ to be energy
independent, but polarization dependent with
$\Gamma^{L}_{\uparrow}=\Gamma^{R}_{\uparrow}=(1+P)\Gamma_{0}$,
$\Gamma^{L}_{\downarrow}=\Gamma^{R}_{\downarrow}=(1-P)\Gamma_{0}$ for P alignment,
while
$\Gamma^{L}_{\uparrow}=\Gamma^{R}_{\downarrow}=(1+P)\Gamma_{0}$,
$\Gamma^{L}_{\downarrow}=\Gamma^{R}_{\uparrow}=(1-P)\Gamma_{0}$ for AP alignment,
where $\Gamma_{0}$ describes the tunneling coupling between the dot and the nonmagnetic leads,
and $0\leq p <1$ characterizes the polarization of leads.
In Fig. 1 we plot the zero-temperature differential conductance of spin up (solid line) and spin down (dashed line)
for FM leads with P alignment as a function of bias voltage. One interesting observation
is that the behavior of differential conductance is quite different for small and large $P$ values.
For small $P=0.2$, Fig. 1(a) shows that the zero-bias peak of differential conductance for nonmagnetic
leads~\cite{Meir93} splittes into two peaks even in the absence of magnetic field, and this splitting can be tuned away by 
applying an appropriate magnetic field [Fig. 1(c)] and can be recovered by increasing the magnetic field further
[Fig. 1(e)]. This result agrees well with that in Ref. ~\cite{Martinek02}, which is due to the splitting of the
dot levels renormalized by the spin-dependent interacting self-energy $\widetilde{\epsilon }_{d\sigma }$
satisfying the self-consistent equation 
$\widetilde{\epsilon }_{d\sigma }=\epsilon_{d\sigma }+\mathop{\rm Re}
\Sigma _{1\sigma }^r\left( \widetilde{\epsilon }_{d\sigma },\widetilde{\epsilon }_{d
\overline{\sigma } } \right)$. For large $P=0.8$ [Figs. 2 (b,d,f)], the behavior of spin up is similar to that
of small $P=0.2$, which shows one peak at approriate value of magnetic field and two peaks when the field is away
from the appropriate value. However for the spin down configuration, the differential conductance shows a small
"flat" near $eV=0$ when the magnetic field is smaller than the appropriate value, while it shows a "dip" at
$eV=0$ when the field is at the appropriate value, and finally it shows a large maximum when the field increases 
further. This interesting behavior can be understood with the help of DOS for spin down 
near the zero bias voltage (shown as
insects of Figs. 1(b,d,f)). The DOS for spin down configuration shows a dip at $\omega=
\widetilde{\epsilon }_{d\downarrow }-\widetilde{\epsilon }_{d\uparrow }$ because 
$\mathop{\rm Re}
\Sigma _{1\downarrow }^r\left( \omega \right)$ grows logarithmically at this value, and the dip position
can be tuned by magnetic field around the appropriate value corresponding
to the single peak for spin up configuration.
In Fig. 2 we plot the differential conductance for FM leads with AP alignment. In this case, the behavior
of differential conductance looks similar for spin up and spin down configurations, which has one peak in the
absence of magnetic field [Fig. 2(a,b)] and two shifted peaks in the presence of magnetic field [Fig. 2(c,d)].
The small difference between the up and down spin configurations is due to the small (compared to the P
alignment case) splitting of renormalized dot levels for different spin configurations at finite voltage
(as shown in the inset of Fig. 2(b)).

\begin{figure}
\includegraphics[width=8.5cm]{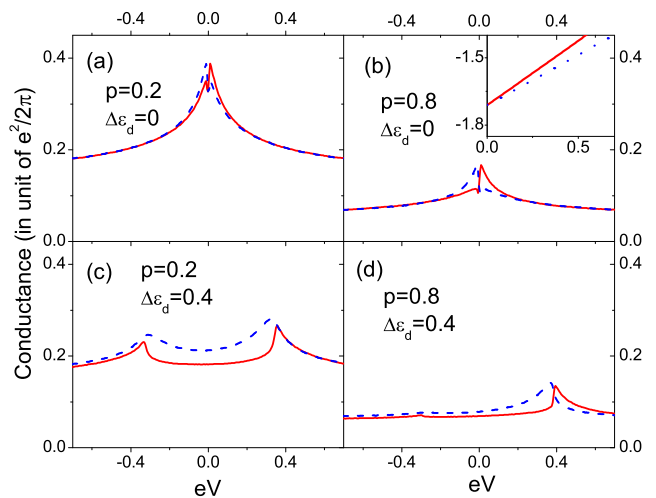}
\caption{\label{fig02} The differential conductance for FM leads with AP alignment. }
\end{figure}

\begin{figure}
\includegraphics[width=8.5cm]{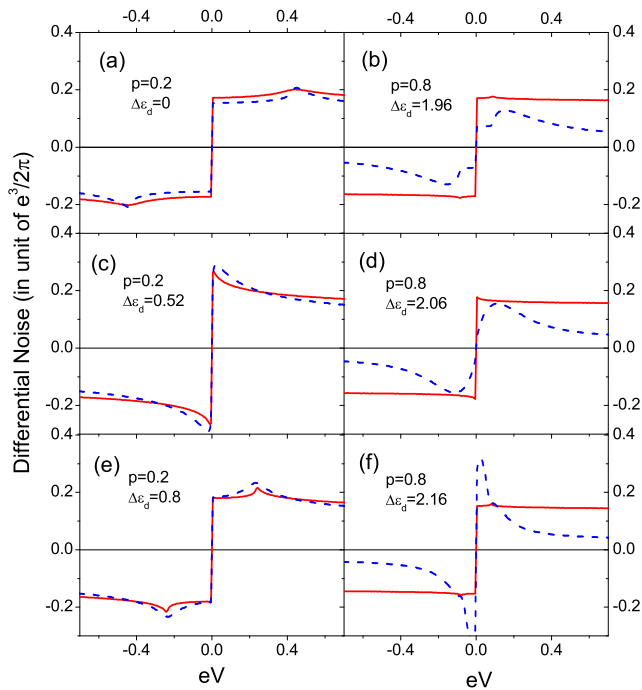}
\caption{\label{fig03} The differential shot noise for FM leads with P alignment.}
\end{figure}

In Figs. 3 and 4, we plot the corresponding differential shot noise for FM leads with P and
AP alignments as a function of the bias voltage. For P alignment case, the differential
shot noise Fiq. 3 also shows the different behavior for spin 
up and down configurations in large
$P$ values, while similar behavior for small $P$ value. 
However, for AP alignment case, the differential shot noise Fig. 4 shows the similar
behavior for different spin configurations. Since the transmission probability corresponding to the
conductance peaks (as shown in Figs. 1 and 2) is small, the differential noise is approximately
proportional to the conductance and only shown one single peak rather than two peaks.

\begin{figure}
\includegraphics[width=8.5cm]{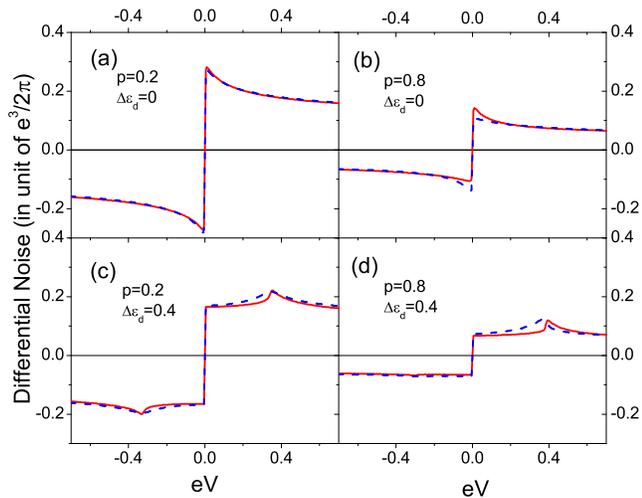}%
\caption{\label{fig04} The differential shot noise for FM leads with AP alignment. }
\end{figure}

To summarize, using the Keldysh nonequilibrium Green function technique, we have studied the
current and shot noise spectroscopy of a single dot with Coulomb interaction coupled to FM leads
with P and AP polarization alignments. We have shown that the lead alignments affect both the current and 
current fluctuations.
For large polarization value, the spin up and spin down configurations have the different behavior in
the differential conductance and shot noise as a function of bias voltage in P alignment case.
While the differential conductance and shot noise show similar behavior for different spin
configurations in P alignment with small polarization value and in AP alignment case with any polarization
value. The derived current and shot noise formulism can be applied to more complicated system.
Work along this line is still in progress.

{\bf Acknowledgments}: We thank Jian-Xin Zhu for many stimulating discussions
which lead to this work. R. L. thanks Xiao-Bing Wang,
M. Sindel, L. Borda, and J. von Delft for useful discussions.

\end{document}